\documentclass[preprint,12pt,authoryear]{elsarticle}

\usepackage[margin=1in]{geometry}
\usepackage{amsmath}
\usepackage{array}
\usepackage{booktabs}
\usepackage{enumitem}
\usepackage{graphicx}
\usepackage[percent]{overpic}
\usepackage{tabularx}
\usepackage{xurl}
\usepackage[hidelinks]{hyperref}
\usepackage[nameinlink,noabbrev]{cleveref}
\usepackage{placeins}

\graphicspath{{figures/}{./}}
\journal{Computers \& Education}

\makeatletter
\def\fnmark[#1]{\def\@fnmark{$\ast$}}
\def\fntext[#1]#2{\g@addto@macro\@fnotes{\def\thefootnote{$\ast$}\footnotetext{#2}}}
\def\corref#1{\def\@corref{$\dagger$}}
\def\cortext[#1]#2{\g@addto@macro\@cornotes{\def\thefootnote{$\dagger$}\footnotetext{#2}}}
\makeatother

\myfooter[L]{Preprint}

\begin{document}

\begin{frontmatter}

\title{Generative AI and Extended Reality in Collaborative Architectural Design Education: An Exploratory Studio Study}

\author[inst1]{Yao Xiao\fnref{equal}}
\author[inst2,inst3]{Max Chen\fnref{equal}}
\author[inst2]{Yichen Li}
\author[inst1]{Nathaniel Powers}
\author[inst1]{Maxwell Wiesenfeld}
\author[inst2]{Gillian Smith}
\author[inst1]{Soroush Farzin}
\author[inst1]{Shichao Liu\corref{cor1}}

\fntext[equal]{These authors contributed equally to this work.}
\cortext[cor1]{Corresponding author.}
\ead{sliu8@wpi.edu}

\affiliation[inst1]{
  organization={Department of Civil, Environmental, and Architectural Engineering, Worcester Polytechnic Institute},
  city={Worcester},
  state={MA},
  country={USA}
}

\affiliation[inst2]{
  organization={Interactive Media and Game Development, Worcester Polytechnic Institute},
  country={USA}
}

\affiliation[inst3]{
  organization={Communication and Media Arts, Western Connecticut State University},
  country={USA}
}

\begin{abstract}
Architectural design education relies heavily on visual ideation and representation to support collaborative learning in studio environments. Recent advances in generative artificial intelligence (GenAI) and extended reality (XR) offer new opportunities for rapid idea exploration and immersive spatial visualization. This exploratory mixed-methods classroom study investigated how GenAI-assisted multi-user XR influenced collaborative architectural conceptual design. We developed GenARch, a pipeline that integrates GenAI-based visual generation with collaborative XR environments, and deployed it in an undergraduate architectural design studio. Twenty-seven students formed seven self-selected design teams; four teams incorporated GenARch into their usual course workflow to support collaborative ideation and visualization, while three teams continued the same course workflow without GenARch. Pre- and post-intervention surveys assessed design self-efficacy, attitudes toward collaborative learning, and teamwork; a seven-member panel evaluated team design presentations; and GenARch teams participated in group interviews. The quantitative results showed larger relative declines in confidence and outcome expectancy for the GenARch condition and a positive difference-in-differences estimate for perceived conflict management, while panel-rated presentation outcomes were not significantly different between conditions. Interviews indicated complementary roles for the technologies: GenAI supported idea externalization and visual reference generation, whereas XR supported spatial, contextual, and scale-based evaluation. Students also reported challenges related to control, dimensional fidelity, shared attention, and motion comfort. These findings highlight both opportunities and limitations when GenAI and XR are incorporated into collaborative design education.
\end{abstract}


\end{frontmatter}

\section{Introduction}

Architectural design requires close collaboration among architects, consultants, contractors, and other stakeholders. Design teams must integrate different perspectives, exchange feedback, and make collective decisions throughout the design process. Architectural conceptual design involves visual ideation, during which abstract thoughts or concepts are translated into more concrete visual forms \citep{safin_design_2016}. These representations are then shared with collaborators and other stakeholders for critique and feedback \citep{dorta_new_2016,scheerlinck_role_2017}. In architectural design education, visual representations similarly provide a shared basis for idea exchange, peer critique, and collective decision-making, making visual communication central to collaborative studio learning.

Advances in generative artificial intelligence (GenAI) and extended reality (XR) present new opportunities for supporting these processes in architectural education \citep{rahimi_generative_2025}. GenAI models can broaden design possibilities and facilitate iterative ideation through rapid visualizations, allowing students and designers to explore conceptual alternatives \citep{karadag_new_2025,lee_impact_2024,yen_search_2025,zhang_optimizing_2026}. Meanwhile, XR technologies, encompassing virtual reality (VR), augmented reality (AR), and mixed reality (MR), provide immersive and interactive environments that can support collaborative design exploration and shared spatial reasoning \citep{jauhiainen_metaverse_2024,raianova_adaptive_2025}. These capabilities are particularly relevant to architectural design, where communication frequently depends on two- and three-dimensional representations.

Recent studies have begun integrating GenAI into XR environments to investigate the complementary advantages of these technologies. Such integration can support visual exploration of geometric form, material alternatives, spatial scale, and contextual qualities \citep{halici_ai-based_2025}. Mixed-initiative and constrained generation approaches also highlight the importance of designer control and autonomy \citep{zhang_vrcopilot_2024}. Together, these studies suggest that GenAI and XR can support different parts of a design workflow: GenAI can help externalize and explore tentative ideas, whereas XR can support spatial inspection and evaluation.

Although GenAI and XR have demonstrated potential for creative exploration, visualization, and collaborative interaction, their combined use in architectural design education remains at an early stage \citep{vallasciani_creaixr_2024}. Research integrating GenAI with collaborative XR workflows remains limited \citep{salazar_rodriguez_exploring_2025}, and many current systems are evaluated through prototype demonstrations or controlled studies rather than in authentic studio courses \citep{rahimi_generative_2025,2025_similar_genaixrinarchi}. Beyond education, recent industry evidence associates stronger returns from AI with its broader integration into organizational processes rather than its use in isolated initiatives \citep{pwc2026ceo}. This perspective raises a related question for education: how can emerging technologies be integrated into existing learning and collaborative design workflows rather than treated as stand-alone tools? Technical affordances alone do not necessarily translate into educational benefits, making it important to examine such integration in authentic learning settings.

To investigate these issues, we developed GenARch, a GenAI-assisted XR pipeline, in consultation with instructors of an undergraduate architectural design studio and deployed it within an ongoing course project. GenARch was incorporated into students' usual course workflow to support collaborative ideation and visualization rather than replace the design tools they normally used. The study used quantitative and qualitative methods as complementary sources of evidence to examine students' learning-related experiences, collaboration, and design processes.

Accordingly, this study addresses the following research questions:

\begin{itemize}

\item \textbf{RQ1:} How does GenAI-assisted XR impact self-efficacy, engagement, and attitudes toward collaborative learning experience in collaborative architectural design education?

\item \textbf{RQ2:} How does GenAI-assisted XR support or constrain teamwork, communication, and shared decision-making during collaborative architectural design?
\item \textbf{RQ3:} How do the roles and perceived value of GenAI and XR vary across different stages of conceptual design, and how are these patterns reflected in students' design processes and design outcomes?
\end{itemize}

These questions focus the study on the educational use of GenAI and XR within an authentic, representation-intensive design workflow.

\section{Related Work}

\subsection{Collaborative learning in architectural design}

Architectural design is a highly collaborative process that requires participants to consider different perspectives, exchange and critique ideas, synthesize information, and make collective decisions. Collaborative learning similarly engages groups of learners in jointly solving problems, completing tasks, or producing shared products \citep{laal_benefits_2012}. Rooted in social perspectives on learning, collaborative learning emphasizes interaction with peers as an important part of knowledge construction \citep{vygotsky_mind_1978}. In architectural design studios, students similarly negotiate design alternatives and integrate individual ideas into a shared proposal.

A central feature of this collaborative process is the use of external representations. Architectural design develops through evolving forms of representation, including sketches, drawings, diagrams, and models \citep{goel_structure_1992}. Students externalize internal design concepts through these representations and use them to communicate, reflect on, and revise their ideas \citep{dorta_design_2008}. In early ideation, sketches and other visual artifacts can function as cognitive artifacts that make tentative ideas available for inspection and discussion \citep{visser_cognitive_2006}. Representations therefore support more than individual visualization; they provide a shared reference through which team members can compare alternatives, provide feedback, and negotiate collective design decisions.

\subsection{3D Generative AI for Architectural Conceptual Design}

GenAI models are increasingly used across the early design pipeline, supporting ideation, rapid prototyping, and sketch- or prompt-guided exploration \citep{tholander_design_2023,jang_generative_2025,li_generative_2025}. Prior studies suggest that GenAI can support concept exploration by enabling students and designers to generate and compare multiple alternatives \citep{ghanbaripour_systematic_2024,jiang2026impactgenerativeaiarchitectural}. Although 3D generation can bridge prompt-level ideation and spatial reasoning, image-based artifacts remain more common in current design workflows, while reliable generation of architecturally usable 3D models remains limited. In a studio-based study in which GenAI was integrated throughout the architectural design workflow, the design-development phase, particularly 3D volumetric detailing, was the least supported, reflecting the immaturity of current tools and the continued need for expert human judgment \citep{alamasi_impact_2026}.

Current 3D-generation approaches range from optimization-driven methods such as DreamFusion to faster feed-forward reconstruction and commercial platforms such as TripoSR \citep{poole_dreamfusion_2022,tochilkin_triposr_2024}. Despite rapid technical progress, architectural applications require more than visual plausibility: designers need geometric fidelity, structural accuracy, consistent scale, and controllability \citep{ling_scenethesis_2025,wang_diffusion_2025}. These constraints make current 3D generation more suitable for early-stage concept exploration than for producing development-ready architectural models. For architectural education, this distinction is important because students must decide when a generated representation is useful for exploration and when direct modeling tools are needed for precise development.

\subsection{Generative XR in Design and Collaboration}

Generative XR integrates GenAI into immersive environments, enabling virtual spaces or content to be created and adapted through human--AI interaction \citep{rezwana_designing_2023}. Unlike XR systems that rely entirely on pre-authored content, generative approaches can support more rapid cycles of scene generation and modification \citep{lv_generative_2023}. Prior work has explored immersive parametric design, generative 3D layout authoring, and language- and vision-guided 3D scene generation \citep{drogemuller_envisioning_2023,zhang_vrcopilot_2024,ling_scenethesis_2025}. These studies demonstrate the growing use of GenAI in immersive design while also highlighting challenges related to controllability, spatial realism, and human creative agency \citep{salazar_rodriguez_exploring_2025}.

The combination of GenAI and XR also has potential for collaborative design. Platforms such as CreAIXR and related multi-user XR systems enable multiple users to create or inspect virtual environments together \citep{vallasciani_creaixr_2024,bussell_generative_2023}. Cross-reality systems have demonstrated benefits for collaboration through shared interfaces and interaction mechanisms \citep{bazargani_beyond_2025}, and AR collaboration has been associated with interaction patterns linked to task performance \citep{you_can_2025}. In this paper, XR is used as an umbrella term that includes VR, AR, and MR; the study does not compare these modalities, but examines the role of shared immersive representations in a Quest~3/Arkio workflow.

Although prior studies establish the potential of GenAI for visual exploration and XR for immersive spatial interaction, their combined use in ongoing architectural design education remains less understood. In particular, relatively little work has examined how students integrate these tools into an authentic course project alongside the conventional resources and modeling tools they already use. This study addresses that gap by examining a classroom deployment of a GenAI--XR workflow for collaborative conceptual design.

\section{Method}

In this work, we developed the GenARch pipeline and deployed it in an undergraduate architectural design studio course. To examine its role in collaborative architectural conceptual design, we adopted a mixed-methods approach combining questionnaire-based self-reports, panel evaluation of team presentations, and semi-structured group interviews.

\subsection{Course Context}

The intervention was deployed in an architectural design studio course that integrates lectures, field observation, and studio-based design development using drawings, physical models, and simulation tools. The course met three times per week for four-hour studio sessions. Before the deployment of GenARch in Week~3, students had completed introductory work in site analysis and architectural conceptual design. The design task asked students to propose a community gathering space, with particular attention to daylight, artificial lighting, and inspiration drawn from nature and everyday experience.

At this stage, students needed support for three core conceptual design activities: generating ideas, translating abstract concepts into 2D or 3D visual form, and evaluating design directions through shared visualization and discussion. These needs motivated the development of GenARch.

\subsection{GenARch Intervention Design}

\begin{figure}[!htbp]
\centering
\includegraphics[width=\linewidth]{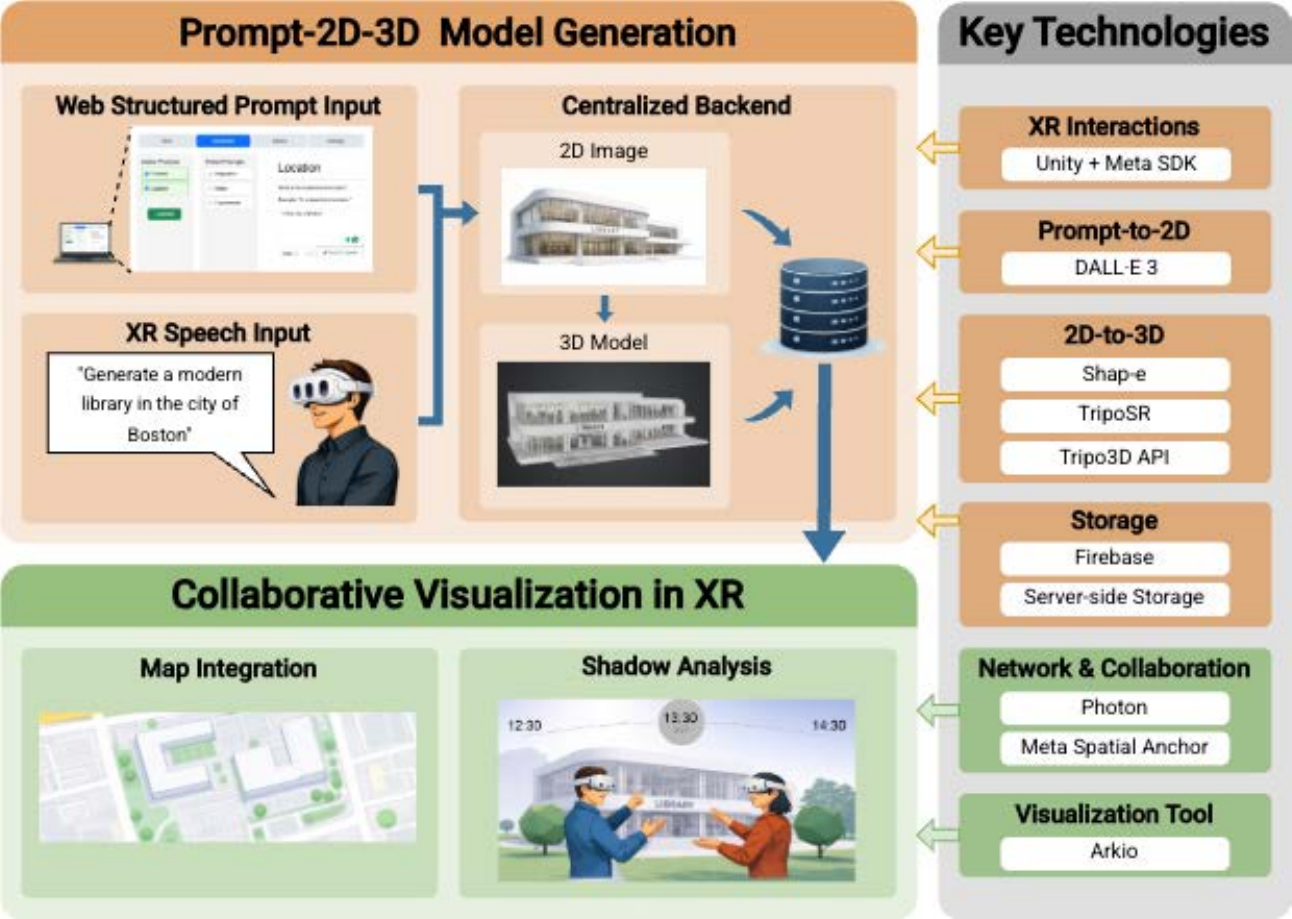}

\caption{GenARch pipeline design and supporting technologies.}
\label{fig:GenARch_pipeline}
\end{figure}

GenARch combines two components (Fig.~\ref{fig:GenARch_pipeline}). The in-house web and XR clients support structured or free-text/speech prompting, 2D image generation, 2D-to-3D conversion, and shared review. Generated assets are stored in a common gallery, allowing students to generate and review ideas without remaining in-headset during processing. Selected assets are then exported to Arkio \citep{arkio}, a commercial multi-user XR platform, for tabletop and full-scale inspection, collaborative manipulation, site/context viewing, and lighting-based shadow analysis.

In this study, GenARch functioned as a \textit{technology probe} \citep{hutchinson_technology_2003} embedded in the existing studio workflow. It was used primarily during conceptual ideation and collaborative visualization, while students continued to use their usual design tools for subsequent design development.

\FloatBarrier

\subsection{Technical Implementation}

The in-house platform used synchronized web and Meta Quest~3 XR clients with a Python/FastAPI backend. DALL$\cdot$E~3 supported text-to-image generation, and Shap-E, TripoSR, and Tripo3D APIs supported 2D-to-3D reconstruction. Photon networking and Meta spatial anchors provided multi-user spatial alignment. These details are reported to make the implemented instructional workflow reproducible while keeping the educational analysis centered on how students used the representations.

\subsection{Participants}

Thirty-five sophomore architectural engineering students were enrolled in the design studio and formed nine self-selected course design teams before study recruitment in September 2025. Two teams declined research participation because at least one member reported severe motion sickness. The remaining seven teams were randomly assigned at the team level: four teams (3--4 students each; $N=16$) incorporated GenARch into their usual course workflow, while three teams ($N=11$) continued the same course workflow without GenARch and served as the Control condition.

A pre-study survey assessed prior GenAI and XR experience on a four-point familiarity scale (1 = never used, 4 = experienced). In the full cohort, 81.5\% reported never or limited familiarity with GenAI and 70.4\% reported never or limited familiarity with XR. Prior use was primarily personal or academic rather than professional or design-oriented. Group comparability was assessed using Fisher's exact tests for categorical variables, Fisher--Freeman--Halton exact tests for familiarity measures, and a Mann--Whitney $U$ test for motion sickness. No significant between-condition differences were found across the assessed pre-study variables ($p>0.05$).

\subsection{Procedure and Measures}

\begin{figure}[!htbp]
\centering
\includegraphics[width=\linewidth]{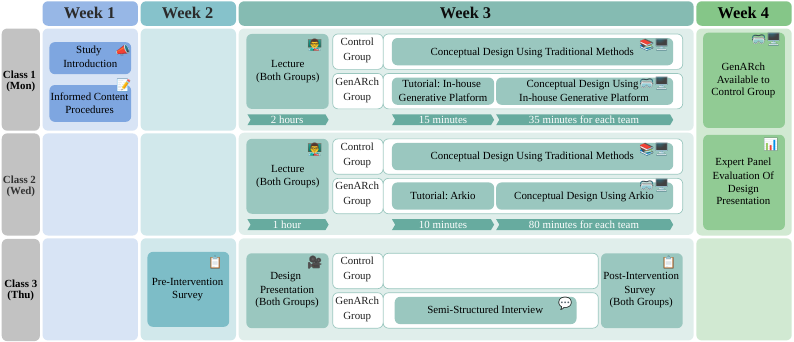}

\caption{Study timeline and intervention procedure for the GenARch and Control conditions.}
\label{fig:course_intervention_timeline}
\end{figure}

The study followed a staged intervention while preserving the course syllabus (Fig.~\ref{fig:course_intervention_timeline}). The protocol was approved by the institutional review board (IRB-24-0780), and all participants provided informed consent. Participation was voluntary and had no effect on course grades. When a student opted out, the whole team was excluded from research data collection. Eight Meta Quest~3 headsets were available, and every participating team member had an individual headset during XR testing.

\FloatBarrier

In Week~1, the research team introduced the study and consent procedures. In Week~2, participating groups completed the pre-intervention measures: prior experience with GenAI and XR \citep{khartabil_exploring_2025}, design self-efficacy adapted from \citet{carberry_measuring_2010}, attitudes toward collaborative learning \citep{vassigh_teaching_2020}, and Team-Q \citep{britton_assessing_2017}. Self-efficacy, collaborative-learning attitudes, and Team-Q were repeated after the intervention. GenARch participants additionally completed the Web-based Learning Tools (WBLT) Evaluation Scale \citep{kay_evaluating_2011} as a post-use measure of perceived learning support, tool quality, and engagement. Table~\ref{tab:measurement} summarizes the constructs and the presentation evaluation.

The GenARch intervention took place in Week~3 across three classes. In Class~1, after the same one-hour course instruction, GenARch teams received a 15-minute tutorial and completed 35 minutes of ideation and asset generation using the web and XR clients. In Class~2, after the same one-hour course instruction, GenARch teams received a 10-minute Arkio tutorial and completed an 80-minute immersive collaborative visualization and evaluation activity. Control teams continued the usual course workflow without GenARch. In Class~3, all groups presented conceptual design outcomes developed through their broader course workflow under the same presentation format. Presentations were recorded for panel evaluation. GenARch teams then completed group interviews (Section~\ref{sec:semi_structured_interview}) and post-intervention surveys. In Week~4, the GenAI/XR tools were made available to the remaining students for educational equity.

Seven raters with architectural-engineering backgrounds evaluated the recorded team presentations using the same 1--5 rubric: two professors and five senior architectural engineering students. The presentation materials did not identify study condition, and the raters could not infer which teams had used GenARch. The rubric assessed design innovation, architectural vision and concept, response to site/climate/lighting, holistic integration, and design process/documentation; the complete rubric is provided in Supplementary Appendix~A. We refer to this group as an \emph{evaluation panel} rather than as a panel of professional experts because five of the seven raters were senior students.

\begin{table}[!htbp]
\caption{\raggedright Measurement instruments and evaluation criteria.}
\label{tab:measurement}
\centering
\footnotesize

{\renewcommand{\tabularxcolumn}[1]{m{#1}}
\begin{tabularx}{\columnwidth}{>{\centering\arraybackslash\hsize=0.5\hsize}X >{\hsize=1.5\hsize}X}
\toprule
Component & Description \\
\midrule

Prior Experience \citep{khartabil_exploring_2025} &
Baseline characteristics and prior experience with architectural design and digital tools. \\

Design Self-Efficacy \citep{carberry_measuring_2010} &
Measures participants’ self-efficacy, motivation, outcome expectancy, and anxiety across key stages of the design process. \\

Attitude of Collaborative Learning \citep{vassigh_teaching_2020} &
Assesses attitudes toward group-based learning, creativity, and peer interaction. \\

Team-Q \citep{britton_assessing_2017} &
Evaluates individual teamwork skills, including contribution to group tasks, facilitation of peer input, team climate, and conflict management. \\

The Web-based Learning Tools (WBLT) \citep{kay_evaluating_2011} &
Post-use measure for GenARch participants capturing perceived learning support, tool quality, and engagement during conceptual design. \\

Presentation Evaluation (Supplementary Appendix~A) &
Final conceptual design outcomes were rated on a 1--5 scale by a seven-member panel comprising two professors and five senior architectural engineering students. The blinded panel used structured criteria assessing innovation, architectural vision, response to site and lighting, holistic integration, and documentation of the design process. \\

\bottomrule
\end{tabularx}
}

\end{table}
\FloatBarrier

\subsection{Statistical Analysis}
\label{sec:stats}

We first compared the GenARch and Control conditions at the pre- and post-intervention time points. Distributional assumptions were assessed with the Shapiro--Wilk test and homogeneity of variance with Levene's test. When both groups met normality and equal-variance assumptions, we used an independent-samples $t$-test; when normality was met but variances were unequal, we used Welch's $t$-test. If normality was violated in either group, we used the Mann--Whitney $U$ test.

For individual pre--post outcomes, difference-in-differences (DiD) regression estimated whether change over time differed between conditions:
\begin{equation}
Y_{it} = \beta_0 + \beta_1 \mathrm{Tool}_i + \beta_2 \mathrm{Post}_t + \beta_3 (\mathrm{Tool}_i \times \mathrm{Post}_t) + \epsilon_{it},
\end{equation}
where $\beta_3$ represents the pre--post change in the GenARch condition relative to Control. DiD was used to compare relative pre--post change while accounting for baseline differences between conditions. Standard errors were clustered at the participant level to account for repeated measures. Given the small number of teams, the inferential results are interpreted cautiously.

For the presentation evaluation, inter-rater reliability across the seven raters was assessed using Kendall's coefficient of concordance (Kendall's $W$) for each criterion \citep{mcleod2005kendall}. The original $W$ values were 0.256 (Design Innovation), 0.234 (Architectural Vision \& Concept), 0.466 (Response to Site, Climate \& Lighting), 0.368 (Holistic Integration), and 0.387 (Design Process \& Documentation). The analysis additionally applied a criterion-specific leave-one-out procedure by omitting one rater at a time and selecting the omission that maximized $\Delta W$. The resulting $W$ values were 0.342, 0.316, 0.553, 0.438, and 0.523, respectively. We report both the original and post-adjustment $W$ values for transparency. Scores were averaged after the leave-one-out procedure, and group results are reported as median [IQR].

Spearman rank correlations were computed between WBLT subscale scores (Learning, Quality, Engagement) and panel-rated outcomes at the team level. All tests were two-tailed with $p=0.05$. Because these associations involved only four GenARch teams, they are interpreted as exploratory.

\subsection{Semi-Structured Interviews}
\label{sec:semi_structured_interview}

Each of the four GenARch teams participated in a semi-structured group interview of approximately 35 minutes. The protocol (Supplementary Appendix~B) addressed learning and engagement, teamwork and communication, design workflow, perceived roles of GenAI and XR, and implementation challenges; probing questions were used when clarification was needed. Audio was transcribed using Otter.ai and manually verified and corrected by a coauthor.

We followed Braun and Clarke's six-phase reflexive thematic analysis \citep{braun_doing_2022,braun_using_2006}. Researchers repeatedly read the verified transcripts, recorded initial observations, and conducted sentence-level inductive coding in Dedoose \citep{dedoose2026}. Codes were clustered into candidate themes and iteratively refined around patterned meanings concerning representational use, collaboration, design-stage needs, implementation frictions, and learner agency. Because the data came from four group interviews and the analysis followed a reflexive thematic approach, themes were developed from patterned meaning rather than frequency thresholds; counts of students endorsing a statement were not treated as prevalence estimates. The qualitative analysis complemented the quantitative patterns by examining how students experienced the intervention and by identifying plausible mechanisms and boundary conditions. The coding framework and coded excerpts are available in the anonymized OSF materials.

\section{Quantitative Results}

\subsection{Self-Perceived Outcomes Across Groups and Time}

Participants completed the pre-survey in $343.4 \pm 129.5$ seconds (two extreme outliers removed) and the post-survey in $275.5 \pm 142.7$ seconds (one extreme outlier removed). The internal consistency of the survey instruments was evaluated using Cronbach's alpha for all multi-item scales and subscales \citep{tavakol_making_2011}. Internal consistency ranged from $\alpha = 0.589$ to $0.965$. Additional experimental details are provided in Supplementary Appendix~C. Figure~\ref{fig:ar_temple} shows examples of architectural design outputs produced using the GenARch pipeline.

\begin{figure}[!htbp]
\centering
\includegraphics[width=\linewidth]{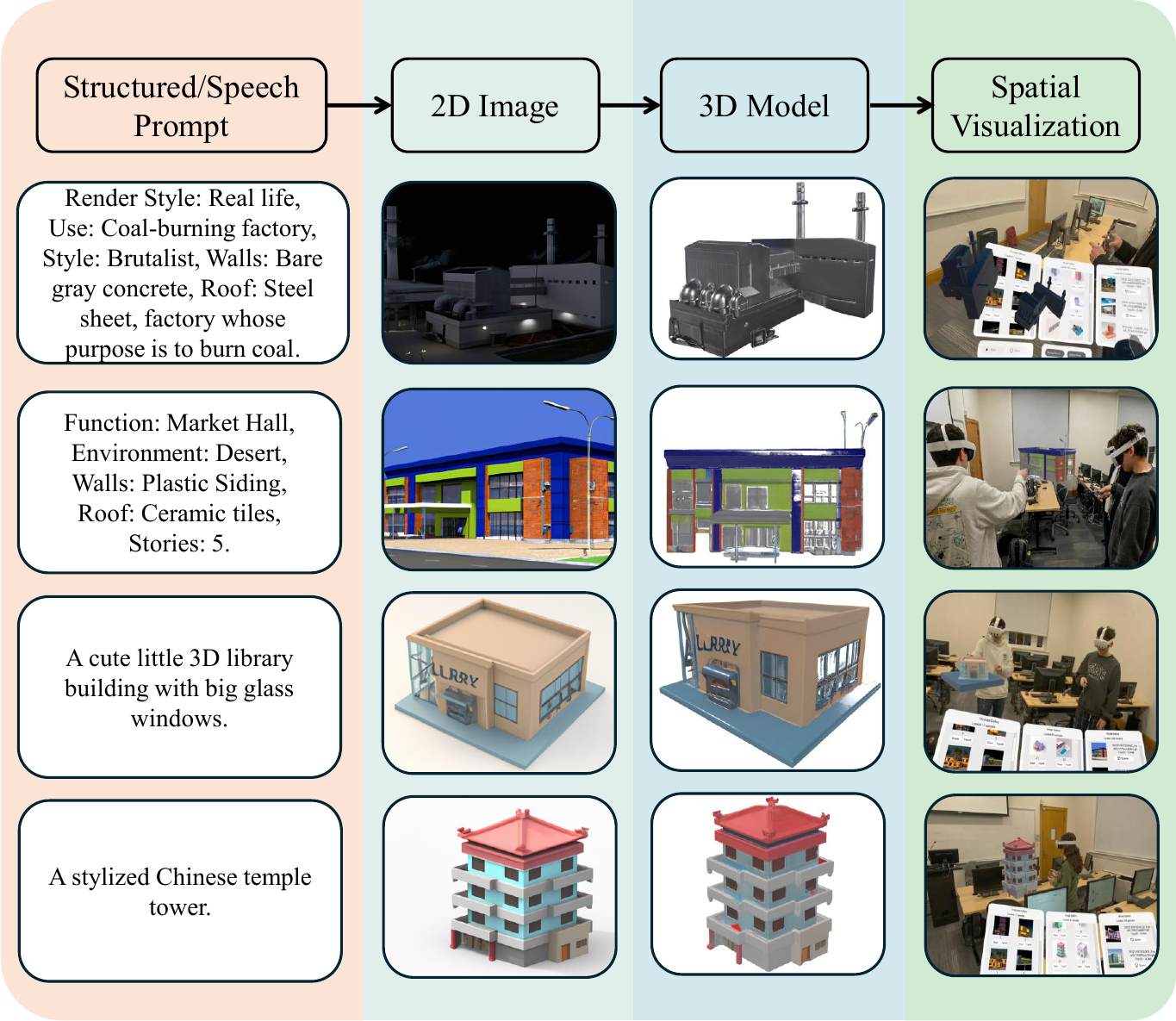}
\caption{Examples of architectural design outputs produced using the GenARch pipeline. A structured or speech prompt describing architectural requirements is first used to generate a 2D image. The generated image is then converted into a 3D model, which is finally visualized in the XR environment for spatial interaction and evaluation.}
\label{fig:ar_temple}
\end{figure}
\FloatBarrier

Pre-intervention comparisons of self-reported outcomes indicated that the GenARch and Control groups were largely comparable across all instruments in \cref{tab:measurement} such as self-efficacy, collaborative learning, and teamwork measures. The only baseline difference was observed for self-efficacy anxiety, with the GenARch group reporting significantly lower anxiety than the Control group ($p = 0.005$).

The DiD estimates (\cref{tab:did}) indicated larger relative pre--post declines in self-efficacy confidence ($\beta = -1.675$, $p = 0.021$) and outcome expectancy ($\beta = -2.088$, $p = 0.002$) for the GenARch condition. For TQ\_D (Manages Potential Conflict), the estimate was positive ($\beta = 0.479$, $p = 0.014$). No significant DiD estimates were observed for attitudes toward collaborative learning, overall teamwork quality, or the remaining teamwork subscales.

\begin{table}[!htbp]
\caption{\raggedright Difference-in-differences estimates for pre--post self-perceived outcomes.}
\label{tab:did}
\centering
\footnotesize

\resizebox{\linewidth}{!}{
\begin{tabular}{lccccccc}
\toprule

 & \multicolumn{2}{c}{GenARch} & \multicolumn{2}{c}{Control} &  &  &  \\
\cmidrule(lr){2-3} \cmidrule(lr){4-5}

Outcomes & Pre-intervention & Post-intervention & Pre-intervention & Post-intervention & DiD & 95\% CI & P value \\

\midrule

SE\_Confidence & $7.578\pm1.247$ & $6.062\pm2.450$ & $7.023\pm1.575$ & $7.182\pm1.199$ & -1.675 & [-3.101, -0.248] & 0.021 \\

SE\_Motivation & $7.516\pm1.665$ & $5.953\pm2.808$ & $6.750\pm2.786$ & $6.727\pm1.690$ & -1.540 & [-3.402, 0.322] & 0.105 \\

SE\_Expectancy & $7.281\pm1.653$ & $5.625\pm2.502$ & $6.909\pm1.206$ & $7.341\pm1.108$ & -2.088 & [-3.414, -0.762] & 0.002 \\

SE\_Anxiety & $3.609\pm2.068$ & $2.734\pm1.997$ & $6.023\pm1.902$ & $5.727\pm2.325$ & -0.580 & [-2.021, 0.862] & 0.431 \\

ACL\_Total & $4.161\pm0.381$ & $4.255\pm0.437$ & $3.985\pm0.320$ & $3.902\pm0.453$ & 0.177 & [-0.161, 0.515] & 0.305 \\

TQ\_Total & $4.444\pm0.440$ & $4.494\pm0.495$ & $4.436\pm0.383$ & $4.155\pm0.582$ & 0.332 & [-0.046, 0.710] & 0.085 \\

TQ\_A & $4.406\pm0.491$ & $4.469\pm0.591$ & $4.545\pm0.522$ & $4.227\pm0.647$ & 0.381 & [-0.128, 0.890] & 0.143 \\

TQ\_B & $4.521\pm0.438$ & $4.521\pm0.596$ & $4.515\pm0.545$ & $4.303\pm0.586$ & 0.212 & [-0.214, 0.638] & 0.329 \\

TQ\_C & $4.656\pm0.473$ & $4.625\pm0.500$ & $4.636\pm0.393$ & $4.364\pm0.710$ & 0.241 & [-0.251, 0.734] & 0.337 \\

TQ\_D & $4.250\pm0.650$ & $4.396\pm0.534$ & $4.152\pm0.689$ & $3.818\pm0.794$ & 0.479 & [0.096, 0.862] & 0.014 \\

\bottomrule
\end{tabular}
}

\vspace{2pt}
\raggedright\footnotesize
SE: Self-efficacy; ACL: Attitude of Collaborative Learning; TQ: Team-Quality; TQ\_A: Contributes to the Team Project; TQ\_B: Facilitates Contributions of Others; TQ\_C: Fosters a Positive Team Climate; TQ\_D: Manages Potential Conflict.
Difference-in-differences (DiD) was estimated using the regression model:
$Y_{it}=\beta_0+\beta_1 Tool_i+\beta_2 Post_t+\beta_3 (Tool_i\times Post_t)+\epsilon_{it}$.
Standard errors were clustered at the participant level. Values are mean $\pm$ SD.

\end{table}
\FloatBarrier

\subsection{Panel Evaluation of Student Design Presentations}

\begin{figure}[!htbp]
\centering
\includegraphics[width=\linewidth]{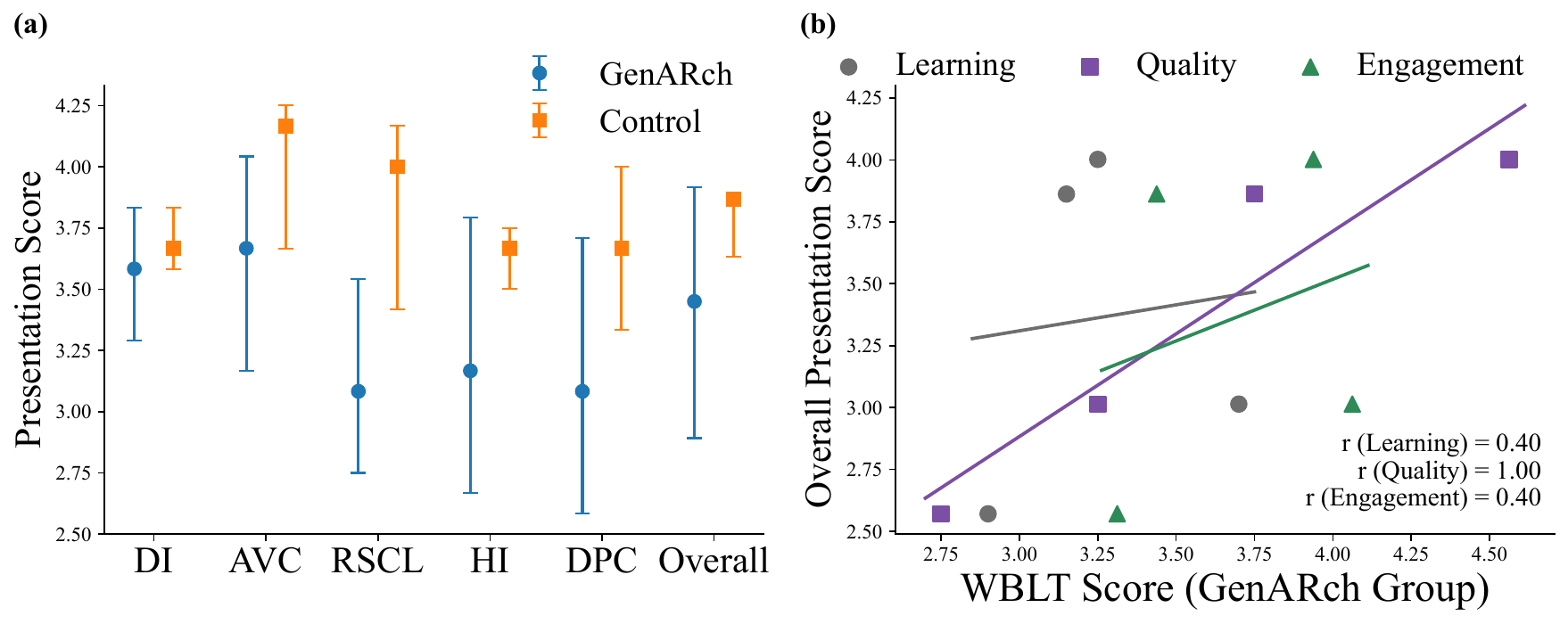}

\caption{Quantitative results. (a) Presentation scores of the GenARch and Control groups. DI: Design Innovation; AVC: Architectural Vision and Concept; RSCL: Response to Site, Climate, and Lighting; HI: Holistic Integration; DPC: Design Process and Documentation. Overall was calculated as each rater's mean across the five criteria and then summarized within condition after the criterion-specific leave-one-out procedure described in Section~\ref{sec:stats}. Values are reported as median [IQR]. (b) Spearman correlations between WBLT subscale scores and panel-rated presentation scores for the four GenARch teams.}
\label{fig:quantitative_results}
\end{figure}

Across the seven team presentations, the GenARch teams showed lower median scores than the Control teams on all five criteria (Fig.~\ref{fig:quantitative_results}a), but the between-condition differences were not statistically significant. The overall presentation score was 3.45 [2.89, 3.92] out of 5 for GenARch and 3.87 [3.63, 3.88] for Control. Greater between-team variability was observed in the GenARch condition. Given the small number of team products and low-to-moderate rater agreement, these comparisons are interpreted descriptively rather than as evidence of a treatment effect.

Figure~\ref{fig:quantitative_results}b summarizes WBLT subscale scores and panel-rated presentation scores for the four GenARch teams. The reported association between perceived tool quality and presentation score was $r = 1.00$ ($p < 0.001$). Because this estimate is based on only four teams, it is retained as an exploratory, hypothesis-generating association and is not used as evidence that perceived tool quality improved design performance.

\FloatBarrier

\section{Qualitative Results}

We found that students considered GenAI text-to-image and text-to-model tools and XR visualization tools useful for different purposes in collaborative architectural design. We first describe what students used or envisioned using GenAI and XR for, then examine how their needs changed across stages of conceptual design, and finally summarize their reflections on creativity, authorship, and control.

\subsection{What students use GenAI and XR for in architectural design}

We identified three closely related subthemes: GenARch as visualization tools, GenARch as a mediator of collaboration and communication, and challenges and implications for future tool design.

\subsubsection{GenARch as visualization tools}

Students used GenAI and XR in complementary ways during the conceptual design process. GenAI primarily supported ideation and visual reference generation, while XR enabled contextual and spatial understanding. In particular, XR helped students visualize the site landscape and perceive architectural elements at realistic scale. By viewing models at full scale in an embodied environment, students reported a clearer sense of proportion and spatial presence compared with screen-based representations. One student explained that, although manipulating very large objects could be inconvenient, ``it was very useful to see the actual sizes of things'' (T3).

In contrast, GenAI tools were mainly used to generate visual references that externalized rough conceptual ideas. Students considered these images useful because they gave the team a more concrete idea before moving into modeling software. At the same time, they wanted the GenAI component to work with the materials already present in their design process, such as site models, sketches, photographs, precedents, or work-in-progress SketchUp geometry. One student imagined combining an existing SketchUp model with AI so that the system could modify the team's own blocks rather than generate an unrelated form from scratch (T2).

\subsubsection{GenARch as a mediator of collaboration and communication}

Students described GenAI-generated images as shared visual anchors for discussion. Teammates could form different mental images from the same verbal description, and generating an image gave them a concrete representation that could be compared, criticized, and revised. As one student described, ``I would generate this image, and say: you guys look at this window, it's kind of cool. And they'd be like, yeah, it looks stupid. We would talk about it'' (T2). In this sense, the value of the image was not only the image itself, but the discussion it enabled.

Students also envisioned XR as a medium for demonstrating work-in-progress or final design proposals to teammates, reviewers, and other stakeholders. However, working with emerging technologies introduced additional coordination demands. Teams sometimes had to troubleshoot the interface while maintaining a shared understanding of the design. Although every participant had an individual headset, simultaneous immersion did not always produce shared attention. One team explained that communication ``didn't really work out'' because members ``were looking at different areas, and we weren't looking at the same thing'' (T4). Students sometimes responded by helping one another, taking turns in particular interactions, or temporarily working individually.

\subsubsection{Challenges, limitations, and implications for future tool design}

Students raised several challenges related to combining GenAI with XR. Some reported motion sickness during XR sessions, and unfamiliar interactions increased the effort required to use the tools. Students also had to learn how to formulate prompts and interpret generated outputs while working on the design task. Small mistakes, such as unintentionally moving objects or triggering generation more than once, were common during early use.

Another challenge was the mismatch between architectural design requirements and the properties of generated models. Students noted that GenAI-generated models did not reliably encode scale, making it difficult to use them directly for spatial reasoning. One student stated, ``I asked it to make a 10 feet wall of 3D model, it wasn't 10 feet'' (T3). Students similarly emphasized the importance of accurate terrain, neighboring buildings, and environmental context in XR. Missing or outdated site information limited their ability to reason about adjacency, lighting, and how a proposed building would sit in the surrounding space.

Students therefore wanted more reliable re-creation and finer control over generation. Prompt history provided some traceability but did not allow outputs to be reproduced or adjusted predictably. Rather than a single black-box generation step, students described workflows in which contextual information could first be established and then smaller components could be generated and edited iteratively. This preference closely reflects the way architectural design develops through repeated refinement rather than one-step generation.

\subsection{Needs for GenAI and XR tools during the various design stages}

Students' reactions indicated that GenAI was generally viewed as more suitable for early exploratory prototyping, while XR was considered more valuable once there was a 3D artifact that could be evaluated spatially. This division also reflected practical constraints: image and model generation could take time, so students often generated assets before or in parallel with immersive inspection.

However, students encountered a paradox when applying GenAI to their projects. During very early ideation, the lack of a concrete direction made it difficult to write prompts that would yield meaningful results. One student explained that the team ``weren't quite at the stage of having ideas for what we wanted the specific building to look like yet,'' so they did not know what to ask the AI to create (T1). At this stage, teams needed to explore broadly, but they did not always have enough conceptual content to express that exploration in text.

Later, when students had more clearly defined concepts, GenAI often struggled to follow the constraints they attempted to encode in prompts. Some students still valued these unexpected outputs as inspiration, while others became frustrated when the system failed to represent an idea they already had in mind. When precision and fidelity became critical, students often preferred direct modeling tools such as SketchUp. In these cases, the desire for control outweighed the benefit of free-form generation. XR followed a different pattern: its value increased when teams had a meaningful 3D representation to inspect for scale, site context, lighting, and spatial relationships.

\subsection{Students' perspectives on the impact of GenAI and XR tools on architectural design}

Students also raised concerns about creativity, authorship, and control during conceptual design. A recurring concern was that generative tools might move from supporting creative work to producing the creative direction itself. Several students framed creativity as a core human contribution to architecture and preferred GenAI to act as a supplementary tool rather than an autonomous generator of architectural form. One student summarized this tension by saying that AI was useful but could also feel like it was ``taking away our creativity'' (T4).

These concerns were connected to understandability and transparency. XR was generally experienced as more tangible and directly controllable, whereas GenAI was more often described as a ``black box.'' Many students preferred workflows that allowed them to incorporate their own sketches, photographs, references, or models because contributing these inputs made the resulting work feel more clearly connected to their own design process. They were more skeptical of generated images or models when the source of inspiration could not be identified or the output could not be reliably reproduced.

Looking ahead, students expressed interest in tools that provide greater control over generated content rather than a single model that must be accepted or rejected as a whole. Editable components, reproducible generation, and better continuity with existing design artifacts could preserve authorship while still allowing GenAI to support exploration. These concerns suggest that the educational role of GenAI in design depends not only on what it can generate, but also on whether students can understand, modify, and take responsibility for the resulting representations.

\section{Discussion and limitations}

This study examined the impact of integrating a structured GenAI--XR workflow into an existing collaborative design course, focusing on students' learning-related experiences, teamwork, and design outcomes. The findings show that the effects of the intervention were not uniform: students identified clear benefits for visual ideation, spatial evaluation, and communication, while the quantitative results did not show consistent improvements across self-efficacy, collaborative learning, teamwork, or panel-rated design outcomes.

\subsection{Integrating GenAI and XR in design education and team collaboration}

Our quantitative analysis indicated a positive DiD estimate for the Team-Q conflict-management subscale, while overall teamwork did not show a significant relative change. The interviews suggest several ways in which GenARch could support team communication. Verbal descriptions often fail to convey an exact mental image, and sketches or prototypes can take time to develop during early ideation. GenAI-generated images provided a faster way to externalize an idea and create a concrete object for discussion. Students could point to a feature, compare interpretations, and negotiate whether an idea should be kept or rejected.

XR provided a different form of common representation by allowing all team members to enter the same spatial model. Every participant had an individual headset during testing, so the primary coordination issue was not headset sharing. Instead, teams sometimes struggled to maintain shared attention while different members viewed or manipulated different areas of the environment. This suggests that multi-user access alone does not guarantee coordinated collaboration. Future collaborative XR environments may benefit from clearer indications of teammates' viewpoints, pointing or focus cues, and mechanisms that help groups deliberately return to the same design element for discussion.

The GenARch condition also showed larger relative declines in confidence and outcome expectancy. These results should be considered together with the implementation challenges described in the interviews. Students were learning unfamiliar interaction methods while also deciding whether generated outputs were appropriate for their design. Motion discomfort, inconsistent scale, latency, and limited control could redirect attention away from the design task. At the same time, direct experience with the tools may have made their limitations more visible to students who entered the study with relatively limited GenAI and XR experience. The study cannot distinguish among these explanations, but the results show why the educational evaluation of emerging technology should consider both intended benefits and the demands introduced by its use.

GenARch did not lead to higher panel-rated presentation scores. This result should be interpreted in light of how GenARch was used within the course. It supported selected ideation and visualization activities, while both conditions continued to use the broader course workflow and conventional modeling tools to develop their final designs. Some students also treated GenARch as exploratory rather than directly integrating every generated artifact into the proposal that was eventually presented. The panel scores therefore reflect the broader course design process, not the quality of a GenARch-generated artifact alone.

\subsection{Users' needs in different phases of creative projects}

The timing of GenAI and XR use was closely related to the changing needs of the design process. Prior work suggests that prompt-based GenAI can support divergent exploration, while more constrained development places greater demands on control and precision \citep{lee_impact_2024}. Students in this study described a similar pattern. At very early stages, they sometimes did not yet know what they wanted to ask the system to generate. As the design developed, prompts became easier to formulate, but students also expected the system to follow more detailed requirements.

This creates a useful tension for design education. GenAI appeared most valuable when students had enough direction to evaluate alternatives but still had room to explore them. XR became more useful after a meaningful 3D artifact existed, when teams could evaluate scale, site context, adjacency, lighting, and spatial presence. When exact geometry and deliberate control became important, students returned to Revit, SketchUp, and other familiar modeling tools. The question is therefore not whether GenAI or XR should be used throughout the project, but when each representation is most useful within the broader design workflow.

The course setting also showed the value of allowing students to experiment with emerging tools without requiring every generated output to become part of the final design. Some teams used GenARch to test ideas that were later abandoned. In a learning context, this exploration is not necessarily unproductive: it can help students understand both the possibilities and the limits of a new technology before deciding how it fits their own design process.

\subsection{Technical considerations for developing GenAI-in-the-loop XR design tools}

At the time of the deployment, image and model generation could take seconds to minutes. The use of both web and XR interfaces allowed students to initiate generation without remaining in the headset throughout the waiting period. This was useful because extended headset use could increase discomfort for some participants. As generation becomes faster, tighter integration between generation and immersive review may become more practical, but speed alone does not address the need for control, scale fidelity, and continuity with existing design artifacts.

Prior work shows that users often rely on embodied gestures to communicate spatial intent when prompting generative systems in VR \citep{aghel_manesh_how_2024}. Our findings add that architectural design students also wanted to specify realistic size and scale, not only where an object should be placed. They also wanted more accurate site context and the ability to build from existing geometry. These requirements are particularly important when generated content moves from visual inspiration to spatial evaluation.

\subsection{Educational Innovation}

This study extends prior research on GenAI and XR by examining their combined educational role within an authentic collaborative design studio rather than through a prototype demonstration or short controlled task. GenARch brought rapid GenAI-supported visual ideation and multi-user XR evaluation into an ongoing course project, allowing the study to examine learning-related perceptions, teamwork, design outcomes, and students' experiences as these technologies became part of an established classroom workflow.

The findings highlight the importance of aligning emerging technologies with the changing needs of the design process. GenAI was particularly useful for externalizing and exploring tentative ideas, whereas XR supported spatial, contextual, and scale-based evaluation after visual or three-dimensional representations had been created. Their educational use was also shaped by prompting, control, dimensional fidelity, shared attention, motion comfort, and continuity across representations. These findings suggest that educational integration should consider when they are introduced, how students move among representations, and how team collaboration and designers' agency are supported throughout the activity. Although architecture is a particularly representation-intensive setting, these considerations may also be relevant to other collaborative and project-based learning domains.

\subsection{Limitations}

This study has several limitations. First, the sample size was relatively small and drawn from a single architectural design studio, which limits statistical power and the generalizability of the findings. Teams were self-selected before recruitment and then assigned as intact teams, so pre-existing team dynamics may have influenced teamwork-related outcomes. Two teams declined participation because of motion sickness, and some participants also experienced discomfort during XR use.

Second, the intervention was conducted over a short period within a course project and involved two unfamiliar technologies. The GenARch teams also received short intervention-specific tutorials in addition to the common course instruction. Students in both conditions could use the resources normally available in the course, so the study evaluates the addition of the structured GenARch workflow rather than isolating the effects of GenAI or XR from all other tools. Third, technical limitations in 2D-to-3D generation, scale fidelity, latency, contextual data, and control were inseparable from the educational experience. Fourth, presentation ratings showed low-to-moderate inter-rater agreement and were based on only seven team products; WBLT--presentation correlations were based on four GenARch teams and should therefore be interpreted cautiously. Finally, the study did not collect systematic observation or reflective-journal data, so the qualitative analysis relies on students' retrospective group-interview accounts.

Future work should first refine GenARch based on student and expert feedback, particularly in interaction usability, dimensional fidelity, controllability, traceability, shared attention, and motion comfort. The improved system should then be evaluated through longitudinal and repeated classroom deployments across multiple design studios to examine how students' experiences, collaboration, and learning-related outcomes evolve with sustained use. Future studies should also include richer process measures, such as structured observations, interaction traces, and reflective journals, together with assessments that more directly capture the ideation, visualization, and collaborative activities supported by GenAI and XR.

\section{Conclusion}

This study examined the integration of GenAI-assisted multi-user XR into an existing collaborative architectural design workflow through GenARch. The quantitative findings showed mixed rather than uniformly positive outcomes during students' initial classroom use of the system. Some learning-related perceptions declined, conflict management improved significantly, and no clear advantages were observed in overall teamwork or design outcomes. The qualitative findings helped explain this pattern by revealing both the complementary value and the practical frictions of the technologies. GenAI supported rapid externalization and exploration of tentative ideas, while XR supported spatial understanding, scale perception, contextual evaluation, and shared review. Students also encountered challenges related to prompting, control, dimensional fidelity, shared attention, and motion comfort, and the perceived value of the technologies changed across different stages of the design process.

The educational implications concern how GenAI and XR are introduced and coordinated within collaborative design learning. The phase-dependent patterns observed in this study suggest that different technologies can support different activities across the design process, with GenAI contributing to rapid visual exploration and XR supporting spatial evaluation and shared review. Their educational value was also closely connected to students' ability to formulate intentions, interpret and control generated outputs, maintain shared attention, and move between representations. These findings highlight the importance of considering timing, scaffolding, representational continuity, and collaborative coordination when emerging technologies are integrated into authentic learning workflows.

\section*{Acknowledgements}
This research was supported by the U.S. National Science Foundation (Award No. 2417102). Any opinions, findings, conclusions, or recommendations expressed in this work are those of the authors and do not necessarily reflect the views of the National Science Foundation.

\section*{Declaration of competing interest}
The authors declare that they have no known competing financial interests or personal relationships that could have appeared to influence the work reported in this paper.

\section*{CRediT authorship contribution statement}
\textbf{Yao Xiao:} Conceptualization, Software, Investigation, Writing -- original draft, Visualization, Methodology, Formal analysis, Writing -- review \& editing. \textbf{Max Chen:} Conceptualization, Software, Writing -- original draft, Methodology, Formal analysis, Writing -- review \& editing. \textbf{Yichen Li:} Software, Writing -- original draft. \textbf{Nathaniel Powers:} Conceptualization, Software, Investigation, Writing -- original draft. \textbf{Maxwell Wiesenfeld:} Conceptualization, Software, Investigation, Writing -- original draft. \textbf{Gillian Smith:} Writing -- review \& editing, Supervision. \textbf{Soroush Farzin:} Writing -- review \& editing, Supervision. \textbf{Shichao Liu:} Writing -- review \& editing, Supervision, Funding acquisition.

\section*{Supplementary materials}

Supplementary material submitted with this article contains the presentation evaluation rubric, semi-structured interview protocol, and additional experimental details (Supplementary Appendices A--C). The qualitative coding framework and coded interview excerpts used in the thematic analysis are available through the anonymized OSF repository at \url{https://osf.io/dzjc5/overview?view_only=b744989d1db1443fb9b44a9d1cd5efab}.

\section*{Data availability}

Anonymized supporting materials are available through the OSF repository above. Public sharing of additional participant-level data is restricted to protect participant confidentiality. Qualified researchers may contact the authors regarding access where permitted by the approved ethics protocol.

\section*{Declaration of generative AI and AI-assisted technologies in the manuscript preparation process}

During the preparation of this work, the authors used ChatGPT (OpenAI) to assist with language editing, manuscript organization, and clarity. After using this tool, the authors reviewed and edited the content as needed and take full responsibility for the content of the publication.

\bibliographystyle{elsarticle-harv}

\bibliography{reference_cae_checked}

\end{document}